# Exoplanets beyond the Conservative Habitable Zone: I. Habitability

Amri Wandel[1]

[1] Racah Inst. Of Physics, The Hebrew University of Jerusalem

Corresponding author, e-mail: amri@huji.ac.il

**Abstract**

The Habitable Zone (HZ) is defined by the possibility of sustaining liquid water on a planetary surface. In the Solar System, the HZ for a conservative climate model extends approximately between the orbits of Earth and Mars. We elaborate on earlier HZ models and apply an analytical climate model of the temperature distribution on tidally-locked planets to extend the HZ. We show that planets orbiting M- and K-dwarf stars may maintain liquid water on their night side, significantly closer to their host star than the inner border of the conservative HZ. We calculate the extended borders of the HZ in the flux–effective temperature diagram. This extension may explain the presence of water vapor and other volatile gases in the transmission spectra of warm Super-Earth-sized exoplanets closely orbiting M dwarfs, recently detected by JWST. We also mention the HZ extension outwards, due to subglacial liquid water in the form of intra-glacial lakes or subglacial melting.

**Keywords**: Habitable zone, Exoplanets, M dwarf stars, Greenhouse effect, planetary climates, Habitable planets, Astrobiology, Exoplanet atmospheres, Tidal radius

1. **Introduction**

The Habitable Zone (HZ) is defined as the circumstellar region where a planet can maintain liquid water on the surface, given the stellar irradiation, or instellation onto the planet, and the atmospheric composition. This concept is rooted in the principle that liquid water is necessary for biochemical processes essential to life, although other factors such as chemical energy sources, elemental diversity, and long-term environmental stability are also important (Chyba & Hand, 2005; Cockell et al., 2016). For the purposes of this study, however, we focus solely on the presence of liquid water as the defining criterion for habitability.

The energy required to keep the water liquid can come from the host star, or, in the case of planets further away from their host star than the outer HZ boundary, from geothermal heat.

Conservatively, the inner boundary of the HZ is determined by the point at which water would evaporate due to excessive heat even before the boiling point is reached, (called the Moist Greenhouse limit, Kasting et al. 1993). The outer HZ boundary is determined by water freezing. Wider empirical HZ boundaries are obtained by assuming that early Mars and recent Venus had liquid water (Kopparapu et al. 2013). These HZ boundaries were calculated with 1D and 3D climate models (Yang et al. 2014; Ramirez and Kaltenegger 2014; Turbet et al. 2016; Kopparapu et al. 2016).

In the Solar System the HZ (for an atmospheric composition of nitrogen, $CO_2$ and $H_2O$) extends roughly between the orbits of Earth and Mars (in the more conservative approach, e.g. Kastings et al. 1993) and between Venus and Mars in the more "optimistic" approach). For other stars the HZ depends also on the star's luminosity. For Main Sequence stars the luminosity is related to the stellar type or surface temperature, hence the HZ also depends on the stellar type of the host.

The HZ and its boundaries also depend on the climate model, as reviewed, e.g., by Shields (2019). Comprehensive 3D calculations of planetary climate model require time consuming Global Circulation Models (GCM), but for an approximate semi-analytic calculation of the HZ borders, faster and more practical 1D models have often been used (e.g. Kopparapu et al. 2013; Wandel 2018; Koll 2022).

Particularly interesting for surveys of exoplanets are M dwarf stars, as they are by far the most abundant type, and their putative habitable planets are easier to detect because of their shorter periods. They are also easier for detecting biosignatures, because of the smaller size of the host star and hence the deeper transit signal for a given planet size.

Potentially habitable exoplanets, of K and M dwarf stars, are expected to be tidally-locked, with one hemisphere permanently facing the star. Initially, this configuration raised concerns about extreme temperature gradients and atmospheric collapse on the dark side (Joshi et al., 1997). However, 3D climate models have demonstrated that given a sufficient atmospheric pressure, or the presence of an ocean, efficient heat redistribution between the day and night sides can stabilize temperatures and maintain habitable conditions (Yang et al., 2014; Leconte et al., 2013; Kopparapu et al., 2017). These results suggest that for tidally-locked planets, the inner edge of the HZ may lie closer to the star than in the case of rapidly rotating planets. For example, Yang et al. (2013) showed that strong substellar cloud formation on tidally-locked planets can increase planetary albedo (here and hereafter this means Bond albedo), allowing their night side to remain cool and stable, even under high stellar fluxes—effectively pushing the inner boundary of the HZ inwards. Moreover, Kopparapu et al. (2016) recalibrated inner HZ limits using such cloud feedback mechanisms, significantly expanding the potential for stable surface liquid water on planets orbiting low-mass stars. These findings challenge earlier assumptions and suggest that tidally-locked planets may not only be common but also viable candidates for habitability, especially in the extended inner regions of the HZ around K and M dwarfs.

Other concerns raised about the habitability of planets orbiting M dwarfs include energetic outbursts in the early evolutionary stages. Such intense X- and UV-radiation bursts and coronal mass eruptions may disintegrate water molecules and erode the atmosphere (e.g. Sanz-Forcada et al. 2011; Luger and Barnes 2015; Tilley et al. 2019). While super-Earth and larger planets are likely to have a thick H/He atmosphere that may survive the early active evolutionary phase of their M dwarf host, the survival of water and atmosphere on smaller, Earth-sized planets was considered less certain. However, some works argue that an atmosphere and water may survive (e.g. Yang et al. 2014; Shields 2016; Wandel and Gale 2020). An additional path to habitability of such planets was proposed by Wandel (2023a): sub-glacial liquid water could survive and replenish the water in the atmosphere once the M dwarf host has calmed down: the ice cover may not be hermetic. Cracks can form and pressurized water could be pushed up to the surface. In addition, surface ice may evaporate if more heat is transported from the day side.

In section 2 we derive approximate scaling relations between the spatial (geometrical) HZ boundaries and the host star mass and temperature, for Main Sequence stars. In Section 3 we calculate new expressions for the extended HZ boundaries for tidally-locked planets, in terms of a global heat-transport fraction. In Section 4 we parametrize the atmospheric greenhouse effect and its impact on the habitability of locked planets in terms of the heating parameter. Finally, in Section 5 we consider the implications of these results for biosignature observations.

## 2. The Habitable-Zone Borders

The HZ depends on the luminosity of the host star, as a planet's equilibrium temperature depends on the stellar irradiation, or instellation. For Main-Sequence host stars, which are the great majority of all stars and the most relevant for the evolution of life, the luminosity is related to the host's stellar type or effective temperature. Consequently, the location and width of the HZ of Main-Sequence stars depend on the stellar type of the host. An approximate expression describing the HZ boundaries can be derived using the relation between the stellar mass and luminosity, $L \sim M^{3.5}$ (e.g., Carroll 2017).

Combining this with the relation between the planet's distance from the host star, $d$, stellar radiative flux $S$, and luminosity, $L$, gives

$$d = (L/L_\odot)^{1/2} (S/S_E)^{-1/2} \text{ AU,} \qquad \text{eq. 1}$$

where the flux is normalized to Earth ($S/S_E$) and the luminosity is in solar units. Using eq. 1 and the Recent Venus and Early Mars HZ-boundaries in the Solar System, $0.75\text{AU} < d_{HZ} < 1.7\text{AU}$, the respective boundaries of the HZ of most Main-Sequence host stars can be written approximately as

$$0.75 \text{ AU} < d_{HZ} / (M/M_\odot)^{1.75} < 1.7 \text{ AU} \qquad \text{eq. 2}$$

This relation is approximately represented by the diagonal orange stripe in Fig. 1. Note that the dashed line marking the center of the HZ has a slope of 1.65, close to the power of $M$ in eq. 2 (the slight difference is due to the power law approximation used for the mass-luminosity relation).

Alternatively, the dependence of the HZ boundaries on the host stellar type may be expressed in terms of the stellar effective temperature, rather than mass, using the relation between effective temperature and luminosity of Main Sequence stars. For a wide range of stellar effective temperatures, *3000K < $T_{eff}$ < 15,000K*, the relation can be approximated by $L \sim T_{eff}^{7}$ (Carroll 2017; Wandel and Gale 2025). Together with eq. 1 this relation gives

$$0.75 \text{ AU} < d_{HZ} / (T_{eff}/T_\odot)^{3.5} < 1.7 \text{ AU}, \qquad \text{eq. 3}$$

where $T_\odot$ is the solar effective temperature.

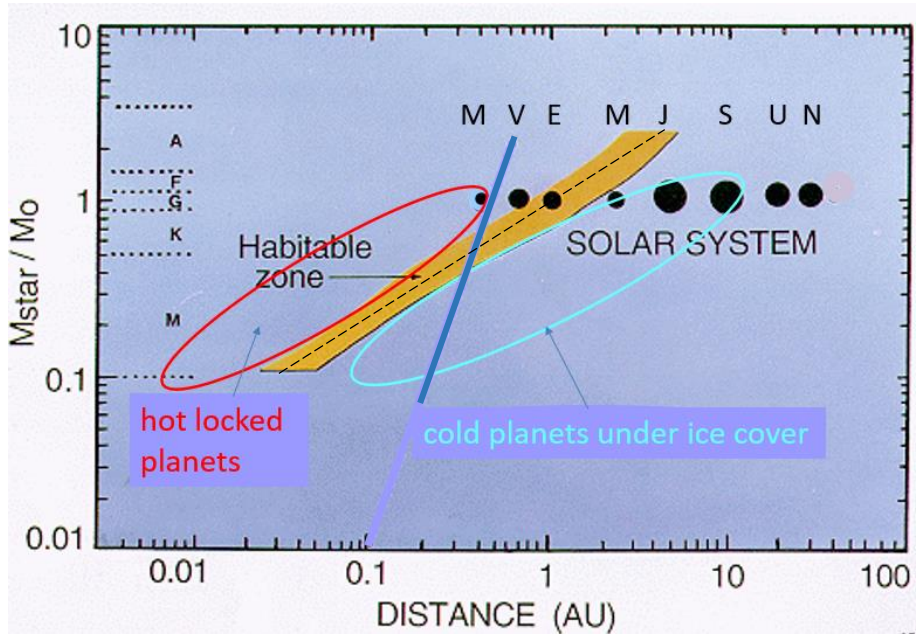

Fig. 1. Extensions to the HZ (the orange stripe) as a function of distance between the planet and its host star (horizontal axis), and the mass of the Main Sequence host (vertical axis). The blue diagonal line denotes the tidal locking radius. The light-blue ellipse represents the outwards extension, while the red one – the extension inwards, for tidally locked planets. (Modified from Fig. 16 of Kasting et al. 1993).

We consider extensions to the Conservative Habitable Zone (CHZ), defined by the above boundaries beyond the inner border of the CHZ (close to the host star) and outwards, beyond its outer boundary (Fig. 1).

A significant extension of the inner HZ boundary can be obtained by the large surface temperature gradient on tidally-locked planets. This planetary climate configuration can support liquid water on the surface (Wandel and Gale 2020), or under an ice cover (Wandel 2023b). Here we derive specific expressions for the corresponding HZ boundaries, depending on the planets heat transport.

The inner HZ extension applies mainly to M and K dwarf stars (the red ellipse in Fig. 1). Their night side may have a surface temperature low enough for liquid water, while being sheltered from the intense UV radiation from flares in the early evolutionary stages of M dwarfs. Sufficiently large planets can maintain a magnetosphere, that would shield their atmosphere against erosion by stellar winds and giant coronal mass ejections (e.g. Rodríguez-Mozos and Moya 2019). On the other hand, on locked planets an atmosphere too dense may induce an efficient heat transport from the day side to the night side, reducing the temperature difference between the night and day sides. This may lead to an increased rate of moist greenhouse evaporation of the night side water or ice.

We show that the global heat transport determines the inner boundary of the extended habitability (the blue dashed lines in Figs. 2 and 3). The HZ can be extended inwards also by increasing the albedo due to a permanent cloud layer, such as in Venus (Selsis 2007), which could considerably lower the surface temperature of the planet.

Extended habitability could exist also outwards of the CHZ (that is, beyond the outer HZ-boundary, marked by the blue ellipse in Fig. 1). This extension could be driven by internal ("geothermal") heat, either due to radiogenic elements (as in the case of Earth) or due to tidal heating (as in the case of Europa and Enceladus). As a conservative outer limit for the HZ extended by subglacial basal melting one may use the flux on Trappist-1 g, the coldest one of the exoplanets simulated by Ojha et al. (2022), which is ~25% of the flux received by Earth (*0.25 $S_E$*). This value is roughly consistent with the outer boundary set by the 1D model (the Maximal Greenhouse limit), as well as with the Early Mars hypothesis. A yet lower flux limit inferring a further out boundary of the HZ may be derived from the evidence for an intra-glacial lake in the south pole of Mars (Arnold *et al.* 2022). We assume that the radiative flux received by the Martian *polar region* may be approximated by an average solar inclination angle of <cos(*i*)> = 0.2. This gives a flux limit of *~0.1 $S_E$*, which is denoted in Fig. 2 by the dashed violet line marked "Martian polar lakes".

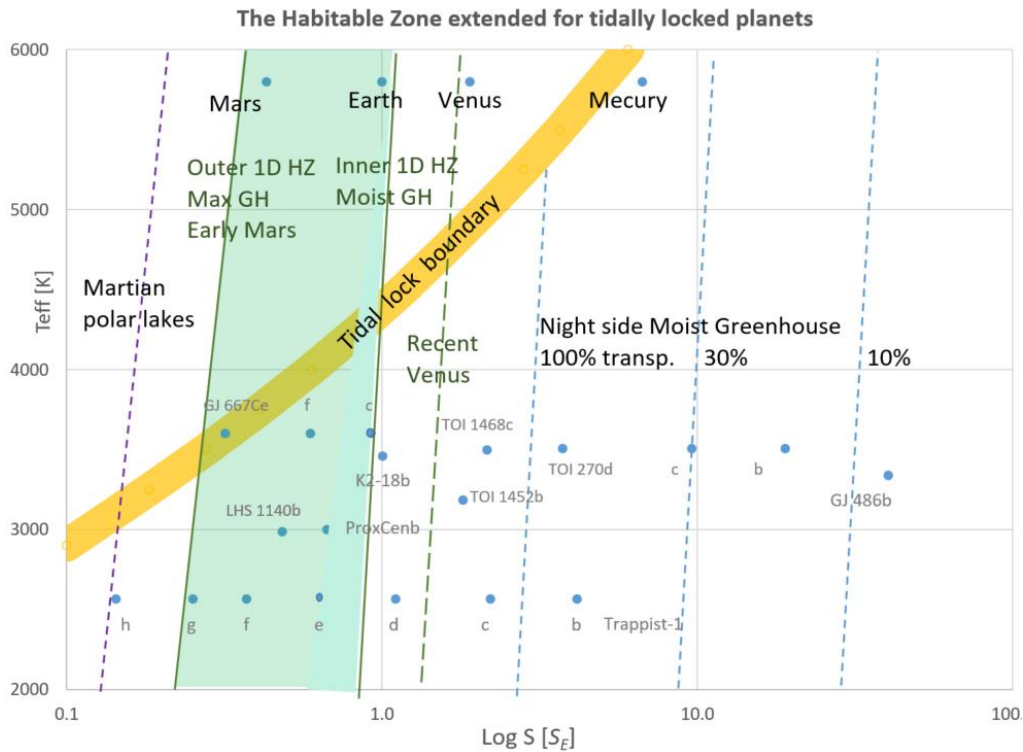

Fig. 2. Boundaries of the Habitable Zone as a function of radiative flux received from the host star, relative to Earth. The green lines and shading denote the conservative HZ boundaries. Dashed blue lines denote the inner extended HZ boundaries. The dashed violet line is the outer HZ extended boundary (Martian polar lake analog). The orange curve marks the tidal locking radius. Blue dots denote solar system planets and some well-known exoplanets orbiting M-dwarfs (modified from Wandel 2023b).

With the relation

$$S/S_E = L/L_\odot (d_{HZ}/AU)^{-2}$$

the HZ boundaries can be expressed in terms of flux, rather than distance from the host star. For example, the Recent Mars and Early Venus boundaries can be approximated by the relationship

$$0.35 < S/S_E < 1.8.$$

These values can be seen in Fig. 2, where the respective curves cross the effective temperature of the Sun, marked by the inner planets of the Solar System (the upper blue dots). However, the HZ-boundaries are not vertical lines (fixed flux) but have a weak dependence on the host stellar type, because the spectral energy distribution, which depends on the stellar type, affects the planetary temperature, i.e. because its impact on the albedo.

The boundaries of the HZ as a function of flux and host star temperature are shown in Fig. 2, and more accurately in Fig. 3. The CHZ boundaries (Kopparapu et al. 2013; Yang et al. 2014) are shown in green. Blue dashed lines correspond to surface or subglacial liquid water on the night side ice on locked planets, calculated using the 1D inner CHZ boundary (Moist Greenhouse) multiplied by the factor derived in the next section (eqs. 6-7). The percentage marks the amount of heat transported from the day side. They are

3. **Habitability of locked planets**

As we have seen in the previous section, planets in the HZ of M dwarf stars are likely to be tidally-locked (Fig. 1). Using a 1D model that takes into account atmospheric heat transport and greenhouse effects (Wandel 2018), we show that surface temperatures allowing liquid water can exist on tidally-locked planets closer to the host star than the inner boundary of the CHZ for a wide range of atmospheric conditions

In order to have a closer and more accurate look at the habitability borders and their extension to locked planets, we have redrawn the inner CHZ-boundary for rapidly rotating planets due to Moist Greenhouse, as well as the outer boundary due to Maximal Greenhouse, using the analytic expressions in Kopparapu et al. (2013). They are described by the solid green curves in Fig. 3. Also show is the Recent Venus inner HZ boundary and the Greenhouse Runaway one (light green dashed curve). The blue dashed curves represent the inner extended HZ boundary for locked planets, assuming 30% of the heating due to the incident stellar radiation at any point on the day side is transported and equally distributed on the whole planet, including the night side. Hence the curve marked "Night ice evaporates" is the moist greenhouse curve shifted to the right by a factor of 1/0.3=3.3. The curve marked "Night Moist Greenhouse" is the moist greenhouse curve shifted by a factor of 9.3 (eq. 6). Similarly, "Night Greenhouse Runaway" is the greenhouse runaway curve (dashed light green) shifted by a factor of 15 (eq. 7).

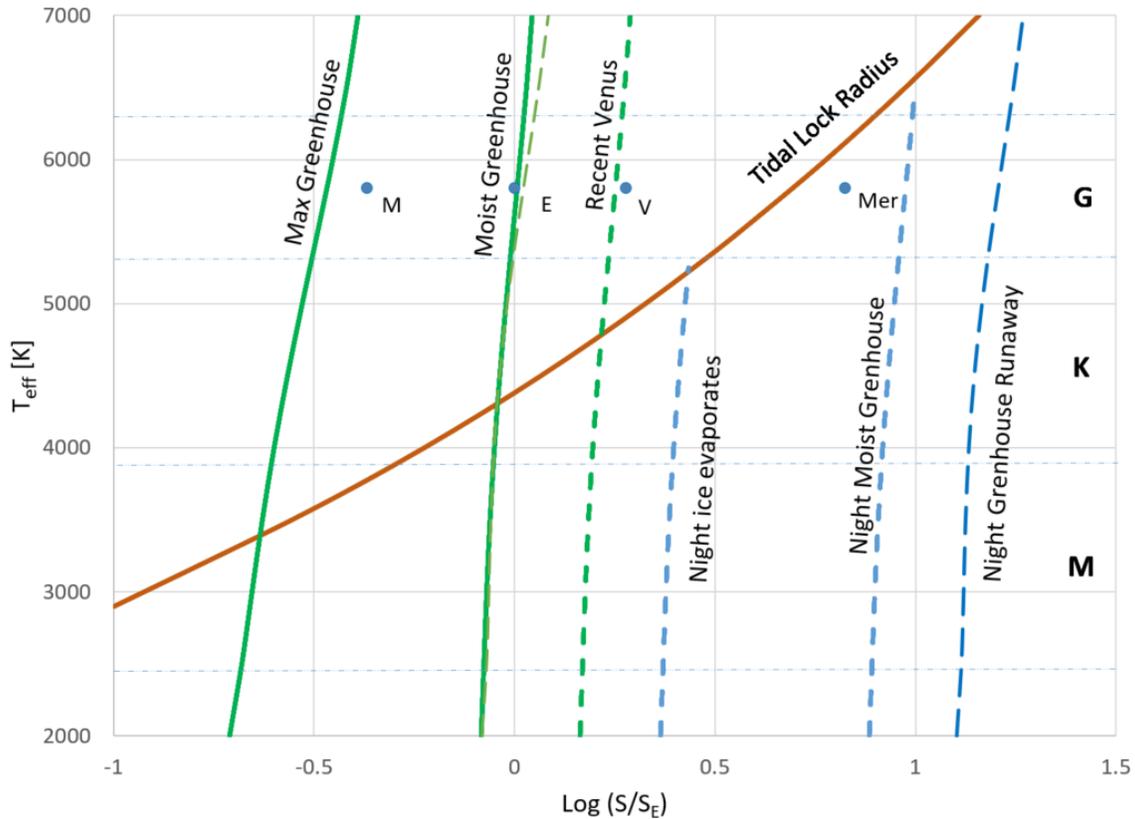

Fig. 3. Boundaries of the Habitable Zone. The horizontal axis marks the radiative flux relative to Earth and the vertical one, the effective surface temperature of the host star. The green lines denote the CHZ boundaries (Kopparapu et al. 2013). Dashed blue lines represent the extended inner HZ boundary considering liquid water or subglacial water on the night side of locked planets. A global heat transport of 30% (transport parameter *f=0.3*) and albedo of 0.3 are assumed. The red curve marks the tidal locking radius. Blue dots denote solar system planets.

We distinguish two scenarios of night side habitability on locked planets:
1) Surface liquid water, when the atmospheric pressure and the temperature at certain latitudes allow liquid water on the surface.
2) Subglacial water, when the day side is too hot for liquid water and the night side too cold, but the global heat transport is low enough to allow ice accumulation on the night side.

In the first case, a global heat transport of (for example) ~30% of the incident radiative heating would sustain liquid water on the night side up to a radiative flux as high as almost ten times that of Earth, *~10$S_E$* (eq. 6).

Also the boundary for night side ice may be derived in this way. The same 1D inner greenhouse boundary of the CHZ (Kopparapu et al. 2013) is assumed to apply to surface ice, and in particular, to night side ice. Under this assumption the CHZ boundary can be shifted to higher instellation, when the heat transport is low enough. E.g. for 30% (*f*=0.3) the extended HZ boundary for night side subglacial water would apply to a flux of ~$3S_E$, (the curve marked by "Night ice evaporates" in Fig. 3).

The atmospheric impact on the surface temperature distribution of a locked planet can be described primarily by two parameters: the heating factor *H* and the heat transport parameter *f* (Wandel 2018). *H* is the effective surface heating by the radiation from the host star, combined with the planet's surface properties (albedo and atmosphere) given by

$$H=(1-A)\, H_g\, S/S_E \qquad \text{eq. (4)}$$

Here *A* is the planet's average albedo, and *S* is the radiative flux from the host star (instellation) and $S_E$ is the solar flux on Earth. $H_g$ is the greenhouse factor, defined as the relative increase of a planet's heating due to the greenhouse effect. $H_g$ is equal to the 4$^{th}$ power of the ratio between the planet's actual surface temperature and the temperature it would have had without an atmosphere, assuming radiative equilibrium. It quantifies the warming caused by atmospheric gases that trap the outgoing long wavelength radiation.

We define the global heat-transport fraction, or the heat transport parameter *f*, as the fraction of the incoming heat transported by advection and distributed over the entire planet. For example, *f*=0.1 means that at any point on the planet surface, 10% of the stellar energy radiated onto the planet is advected and finally distributed equally all over the whole planet, while *f*=1 means that the entire incoming stellar radiation energy is evenly distributed over the whole planet, leading to an isothermal surface.

Of course, the heat transport could depend on many factors, such as rotation period, atmospheric pressure and composition, clouds, wind speed etc. Over a steep temperature gradient, heat dissipation via turbulence may be important. We assume that, to a first approximation, the global heat transport may be represented by a single scalar. In addition to the global heat-transport, Wandel (2018) introduces a local heat transport parameter "*b*", which takes care of smaller scale effects, like turbulence, conductivity of air and ocean etc. However, the local heat transport is typically small compared to the global one and mainly contributes to the form of the transition from dayside high temperatures to the night side low temperatures.

The transport parameter may be estimated for planets in our Solar System (Wandel 2018). For example, for Mars *f*~0.85 while for Mercury *f*~0.003. Orbital resonance (i.e. a synchronously rotating planet like Mercury) gives a more effective heat distribution than in the case of a completely locked planet (Wandel and Gale 2020). It can be approximated by effectively increasing the transport parameter.

Fig. 4 (Wandel 2018) shows the surface temperature distribution as a function of "latitude" (defined as the angular distance from the sub stellar point, on the surface of a locked planet) for various values of the heating factor *H*, defined in eq. 4. Note that for a wide range of *H*, liquid water can exist on part of the planet's surface.

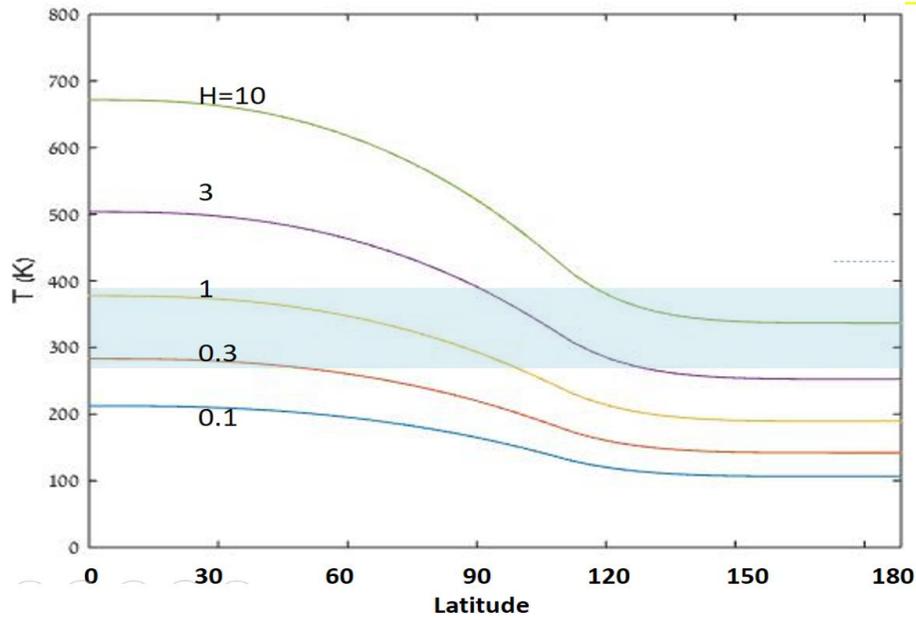

Fig. 4. Surface temperature distribution on a locked planet vs. latitude, as measured from the sub-stellar point (Wandel 2018). Curves are marked by the value of the heating factor *H*. A global heat transport *f*=0.2 is assumed. The light blue stripe marks the liquid water regime (for a 1 bar atmospheric pressure).

The closest distance from the host star at which liquid water would be stable on the surface of a locked planed can be calculated by applying the Moist Greenhouse limit to the night side. The temperature of the off-stellar point (latitude 180 in Fig. 4) is given by (eq. 2b in Wandel 2018)

$$T_{min}=278\ [f(1-A)\ H_g\ S/S_E]^{1/4}\ \text{K}. \qquad \text{eq. (5)}$$

As realistic parameter values we assume an albedo of rock (*A*~0.3), a Moist Greenhouse limit of $T_{min}$~330K (Wolf et al. 2017), and a negligible abundance of greenhouse gases ($H_g$~1). Then eq. 5 gives

$$S/S_E<2.8f^{-1}. \qquad \text{eq. (6)}$$

For instance, the inner HZ instellation limit for a locked planet with *f*=0.3 is $S/S_E$<9.3. This boundary is shown in Fig. 2 by the dashed blue curve marked "30% Night side Moist GH" and in Fig. 3 by "Night Moist Greenhouse".

In the case of effective global heat transport (*f*~1) eq. 6 gives a lower flux boundary, as the planet becomes isothermal and the conservative Moist Greenhouse limit applies. Using instead of the Moist Greenhouse limit the Runaway Greenhouse limit, $T_{min}$~373K (for a ~1 bar atmosphere), gives a somewhat higher instellation limit (and hence a more inward extended HZ boundary). Assuming *A*~0.3 and $H_g$~1 as before, eq. 5 gives

$$S/S_E<4.6f^{-1}. \qquad \text{eq. (7)}$$

For *f=0.3 eq. 7* gives $S/S_E<15$, shown by the long-dashed blue curve marked "Night Greenhouse Runaway" in Fig. 3. The dependence of the flux boundary on the host star type is assumed to follow that of the Greenhouse Runaway limit (Kopparapu et al. 2013; light green dashed curve in Fig. 3).

Similar boundaries may be applied also to subglacial liquid water on locked planets. The night-side ice is assumed to evaporate at the Moist Greenhouse flux limit. As in the case of surface liquid water, the flux limit is assumed to $f^{-1}$ times the conservative Moist Greenhouse flux limit. For example, for *f=0.3* the flux limit is 3.3 times the conservative Moist Greenhouse flux limit.

The functional dependence of the flux boundaries on the host stellar type (Fig. 3) is assumed to be the same as that of the corresponding boundaries for rapidly rotating planets (Kopparapu et al. 2013).

If the planet is not completely locked but slowly rotating, e.g. on a synchronous orbit, the off stellar side would warm up typically within a few days (Wandel and Gale 2020). The coefficient in eq. 5 would then be higher, and consequently the corresponding flux limit would be lower.

In the extreme case of rapidly rotating planets (effectively unlocked) this limit would be significantly lower. E.g., for Earth we have *f=0.95* and *(1-A) $H_g$ =1.15* (Wandel 2018 table 1) and eq. 5 (again with $T_{min}$~330K) gives $S/S_E<1.8$, which is close to the Recent Venus boundary.

It is possible to write a simple expression for the planetary equilibrium temperature. Expressing the instellation later in terms of the properties of the host star and the distance of the planet (eq. 1) gives a useful relation between the host star properties (surface temperature $T_s$ and radius $R_s$) and the planet's distance, *d* and surface equilibrium temperature $T_p$ (e.g., Wandel and Gale 2024, Ch. 14):

$$T_p = T_s [(1-A)^{1/2} R_s /2d]^{1/2}. \qquad \text{eq. (8)}$$

Eq. (8) assumes no greenhouse heating (i.e., $H_g=1$). For Locked planets, a similar expression (without the factor 2 in the denominator) gives the temperature at the sub stellar point.

For locked planets the extremal values of the surface temperature (the maximal, sub-stellar temperature and the minimal, night side temperature (latitudes 0 and 180, respectively, in Fig. 4) depend on the distance from the host star, and on the transport parameter *f*. Fig. 5 (Wandel 2024a) shows the range of surface temperature on locked planets is plotted as a function of distance from the host star and for several values of the transport parameter *f*. The host star is assumed to be an M dwarf with a surface temperature of 3,400 K. For example, the green vertical rectangle shows the temperature range on the surface of a planet orbiting at 0.2 AU. Assuming the transport parameter is 0.5 (50% of the heat due to the stellar radiation is equally redistributed over the entire planet), the temperature at the sub stellar point is approximately 310 K, while the temperature at the coldest point on the night side is 60K. Increasing the transport parameter would reduce the temperature difference; for *f=1*, the planet becomes isothermal (black curve in Fig. 5) with *T*~250K for a=0.2 AU.

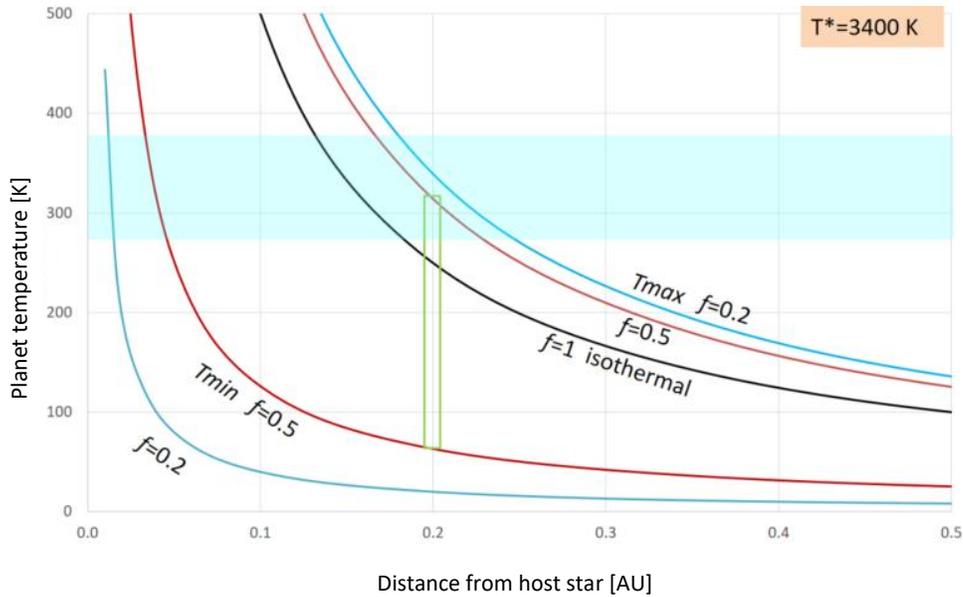

Fig. 5. Maximum (sub-stellar point) and minimum (night-side off-stellar point) temperatures for locked planets orbiting an M-dwarf of T=3400K. Curves are marked by their value of the global heat transport *f*. The light blue stripe marks the liquid water regime (for a 1 bar atmosphere).

**4. The Heating Parameter**

The heating parameter is closely related to the planetary greenhouse factor $H_g$ (eq. 4), determined by the atmospheric density and composition.

For example, let us assume an albedo of rock (*A*~0.3). We can use eq. (4) to demonstrate the relation between the heating factor *H* and the instellation *S*. For an atmosphere with a low greenhouse effect ($H_g$ ~1) we get $H=0.7S/S_E$, while a large greenhouse effect, e.g. $H_g=7$, yields $H$~$5S/S_E$. For comparison, Venus' greenhouse factor is ~26, and with *A*=0.8 eq. 4 gives *H*~10. Note that the definitions in Wandel (2018) are slightly different, e.g. $H_g$ is called $H_{atm}$ etc.

Usually, the HZ is expressed in terms of distance from the host star and the planet's surface temperature, namely, the circumstellar region between the inner HZ edge (water evaporation) and the outer edge (water freezing). This definition does not take into account planetary features such as albedo and greenhouse heating. Following Wandel (2018) we define the habitability range, in terms of the heating. At the lower limit of the habitable range of the heating factor, the temperature on a planet at a given distance would be the freezing point, and at the upper boundary – the evaporation point. The latter could also be applied to ice on the night side of a locked planet with no atmosphere or with a low transport parameter.

This is schematically shown in Fig. 6 (Wandel 2024b). The green region between the solid lines corresponds to surface liquid water at some intermediate latitude on a locked planet with low global heat transport. The blue solid line denotes an "Eye Ball" climate model, with a sub stellar lake and the rest of the planet frozen. The pink solid line denotes an "Inverse Eye Ball" climate, with an off-stellar lake on the night side, the rest of the planet being arid and too hot (in the case of a relatively water-poor planet) or, in the case of a water world, an overall ocean, too hot for life as we know it (*T*>150C) and a massive high pressure atmosphere.

The blue region below the green one denotes subglacial liquid water on planets beyond the outer boundary of the CHZ, where significant heating (e.g. by a high greenhouse effect) is required even in order to maintain subglacial melting (Ojha et al. 2022). The pink region, above the green one, denotes surface or subglacial liquid water on the night side of locked planets. These planets are orbiting their host star closer than the inner boundary of the CHZ, and have a low heat transport that enables relatively low temperatures and surface or subglacial liquid water on the night side. The distance of the planet from the host star is marked on the horizontal axis (assuming a host star luminosity of $10^{-4} L_\odot$). For different luminosities the distances on the horizontal axis scale as $d \sim L^{1/2}$. For example, for a host luminosity of $10^{-2} L_\odot$, the marks on the horizontal axis would be 0.1, 1 and 10 AU instead of 0.01, 0.1 and 1 AU, respectively.

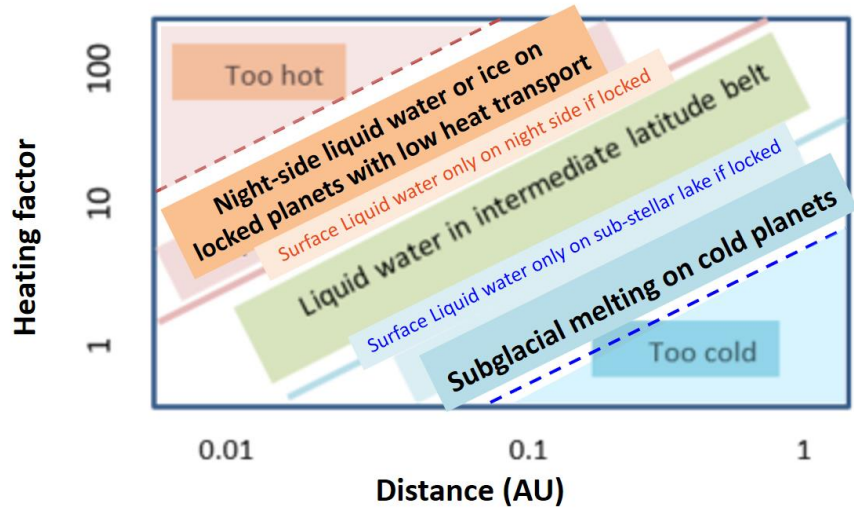

Fig. 6. A schematic presentation of the extended HZ (in terms of the "habitability" heating factor, on the vertical axis) for surface liquid water and subglacial water on locked planets (green, between solid lines) and for extreme conditions (pink and blue, between dashed lines). The distance of the planet from the host star, marked on the horizontal axis, scales as the square root of the host luminosity. The values shown are for $L=10^{-4} L_\odot$.

While Fig. 6 shows the habitable heating factor boundaries as a function of distance from the host star, and for a specific host luminosity, it is possible to draw a analogous "universal" diagram by switching the horizontal axis from distance to flux units (Fig. 7). The green vertical bars indicate the flux on some specific exoplanets (marked in Fig. 2), having various instellations.

For a given global heat transport parameter (*f*) one can estimate the range of the heating factor allowed by habitability, for a planet with a given instellation. This is indicated by the red and blue lines in Fig. 7. For example, the green rectangle shows the *H*-range allowing liquid water (for *f=1*) on part of the surface of K2-18 b, assuming a 1bar atmospheric pressure. Assuming a lower heat transport, e.g. *f=0.2,* the *H*-range is significantly extended, as shown by the orange rectangle for TOI-1468 c.

Planets with high instellation could be too hot for liquid water even on their night side (unless the heat transport is very low), as shown by the blue rectangle for TOI 270 b. In this case, unless $f<0.2$, surface liquid water requires the heating parameter to be less than 1, which can be obtained only with a very high albedo. However, if there is no atmosphere, the effective transport parameter may be very low: e.g. Mercury has $f\sim 0.003$ (Wandel 2018). Of course, surface liquid water cannot exist without an atmosphere; in this case only subglacial water may be considered, and the evaporation boundary would refer to ice rather than to water. The presence of an atmosphere can be inferred from comparing the calculated planetary equilibrium temperature to the temperature observed during a transit (e.g. Zieba et al. 2023).

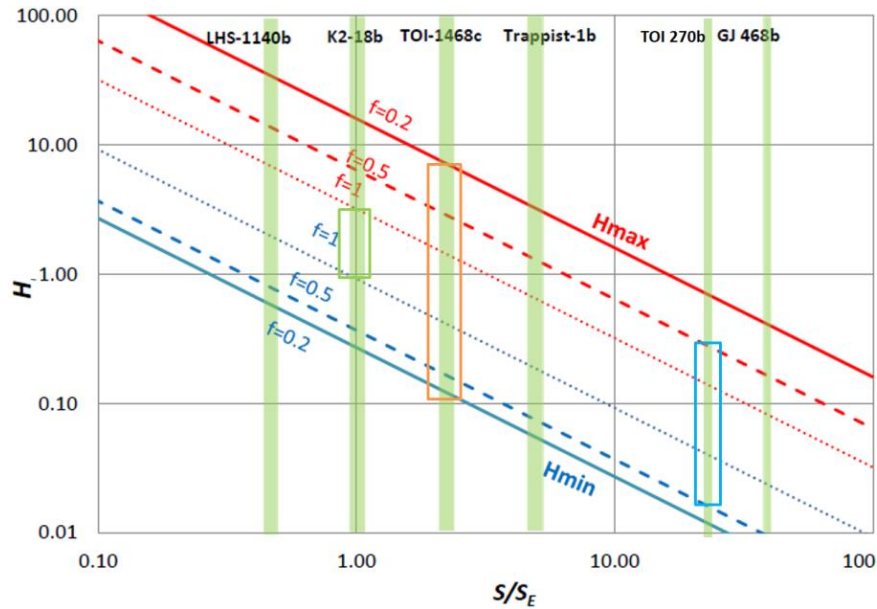

Fig. 7. A "universal diagram" for the habitability range in the Heating factor vs. stellar irradiation onto the planet relative to Earth. The diagonal lines (marked by their value of the global heat transport $f$) indicate maximal (red) and minimal (blue) values of $H$, allowing surface liquid water for a given instellation (cf. Wandel and Tal-Or 2019; Wandel 2024a). The vertical green lines indicate a few well known exoplanets. The vertical columns are examples of the liquid water heating factor range for a few of the planets.

The habitability analyses can be combined with the observable parameters of exoplanets (mass and radius) and with the inferred internal structure to estimate the atmosphere and the habitability (Daspute et al. 2025) to give constraints on the atmospheric composition and properties and its impact on the planetary climate.

### 5. Implications for the search of biosignatures

HZ planets of M dwarfs are easier to detect. Being closer to their host stars they have shorter periods which enable more transits in a given observing season. Planets of M dwarfs also have relatively deeper transits compared to those planets with a similar size orbiting brighter Main

Sequence stars, because M dwarfs have smaller radii. For the same reasons it is easier to obtain transmission spectra of M dwarf planets, with a better signal-to-noise ratio.

It has been suggested that planets orbiting close to M stars, may lose their primordial water due to the host's energetic outbursts in the early evolutionary stages, producing intense X- and UV radiation and stellar eruptions. It was also claimed, that stellar winds could erode the atmospheres of M-dwarf planets. These effects would be even more severe for planets in the inner extended HZ.

In cases of limited data with poor quality transmission spectroscopy, the presence of an atmosphere can be tested by infrared eclipse photometry, looking for atmospheric heat transport (e.g. Koll et al. 2019). Significant heat transport would reduce the day side temperature measured at the secondary eclipse, and enhance the night side temperature observed during the primary eclipse. Failure to detect such an effect can give an upper limit for the thickness of an atmosphere (e.g. Kreidberg et al. 2019 for LHS 3844b, Greene et al. 2023 for Trappist 1 b).

Nevertheless, recently signs of water vapor and volatiles have been detected in JWST transmission spectra of small exoplanets. Some of these exoplanets are closer to their M dwarf hosts than the inner CHZ boundary. In particular, signs for water vapor or volatiles may have been detected on e.g. GJ 486 b (Moran et al. 2023) and TOI 270 d, (Holmberg and Madhusudhan 2024). Water detection on such planets is intriguing, since one would doubt the survival of atmosphere and water under such harsh conditions; even the quiescent radiative flux from the host star on these planets is significantly higher than on Earth. Also the calculated equilibrium surface temperature is closer to that of Venus, than to Earth. Being closer to their host star than the inner boundary of the CHZ, finding volatiles on such planets challenge the CHZ model. If confirmed, this could be an empirical evidence for the existence of surface or subglacial water on the night side, supporting the extended HZ model for tidally-locked planets of M dwarfs.

Extended habitability could exist also outwards of the CHZ (that is beyond the outer HZ-boundary), where subglacial liquid water may exist in geothermally heated rocky exoplanets even with low surface temperatures (Ojha et al. 2022, Wandel 2023a). The geothermal heat can be due to radiogenic elements (as in the case of Earth) or due to tidal heating (as in the case of Europa and Enceladus). A further outer boundary of the HZ may be derived from the evidence for an intra-glacial lake in Mars' south pole (Arnold *et al.* 2022; Wandel 2023b).

Finally, we note that the extended HZ implies a higher occurrence of habitable planets per star (Dressing and Charbonneau 2015; Wandel 2023a). The number of habitable planets within the extended HZ of M and K dwarfs may be bigger by a factor of 50, compared with the CHZ of Sun like host stars (Wandel 2025).

## 6. Conclusions

The concept of Habitable Zone in its conservative definition - liquid water on the surface of rapidly rotating planets – is extended considering tidally-locked planets and night side surface and subsurface water. Earlier results for locked planets are applied to derive extended HZ-boundaries for tidally-locked rocky planets. It is found that locked planets can maintain liquid water on their night side at significantly larger incident flux values, than rapidly rotating planets. The extended flux boundary depends on the global heat transport (or hear redistribution). For instance, with 30% of the heat being globally transported, liquid water could survive on the night side of locked planets orbiting an M dwarf about 3 times closer than the inner boundary of the CHZ. A similar extension applies to subglacial water on the night side of locked planets, which is particularly relevant in the case of no or very thin atmosphere. These results are presented in terms of the habitable range of the heating factor, a dimensionless combination of instellation, atmospheric heating (greenhouse effect) and albedo. This extension of the HZ may explain recent observations, of water vapor and volatiles on small planets with high instellation, orbiting M dwarfs beyond the inner boundary of the CHZ.

**Acknowledgements:** this research has been supported by the Minerva Foundation, at the Center for Studying the Planetary Emergence of Life.

**References**

Arnold N.S., Butcher F.E.G., and Conway, S.J., *et al.* 2022, *Nature Astron.* 6, 1256.
Carrol, B.W. 2017, *An Introd. To Modern Astrophys.,* CUP, (Chapter 13) , ISBN 9781108422161.
Chyba C.F. and Hand, K. P. 2005 Ann. Rev. Ast. & Aphys. 43, 31.
CockeTll C. S., Bush, T., Bryce, C. et al., 2016 Astrobiology, 16 (1), 89
Daspute, M., Wandel, A., Tal-Or, L., Kopparapu R. K., 2025, ApJ. 979,158.
Dressing, C.D. and Charbonneau, D. 2015, Ap.J. 807, 45.
Greene, T.P., Bell, T.J., Ducrot, E., et al. 2023, Nature 618, 39.
Holmberg M., and Madhusudhan, N. 2024 A&A 683, L2.
Joshi M. M., Haberle, R. M., and Reynolds, R. T., 1997 Icarus 129 450.
Kasting J. F., Whitmire D. P. and Reynolds R. T. 1993 Icarus 101, 108.
Kasting J. F., Kopparapu R. K., Ramirez R., and Herman, C.F. 2013 PNAS 111 (35) 12641-12646
Kopparapu R. K., Ramirez R., Kasting J. F. et al. 2013 ApJ. 770, 82.
Kopparapu R.K., Wolf, E.T., Haqq-Mirsa, J., et al. 2016, ApJ. 819, 84.
Kopparapu, R.K., Wolf, E.T., Arney, G., et al., 2017, ApJ. 585, 5.
Koll, D.D.B., Malik, M., Mansfield, M. et al. 2019, ApJ. 886, 140.
Koll, D.D.B. 2022, ApJ. 924, 134.
Kreidberg, L., Koll, D.D.B., Morley, C., et al. 2019, Nature 573, 87.
Leconte, J., Forget, F, Chamay, B., et al., 2013 A&A 554, 69.
Luger, R. and Barnes, R. 2015, Astrobiology, 15, 2.
Menou, K. 2013 ApJ. 774, 51.
Moran, S. E., Stevenson, K. B., Sing, D. K., et al. 2023, ApJ. 948, L11
Ojha L., Troncone B., Buffo J. et al. 2022 Nature Comm. 13, 7521.
Ramirez, R.M., and Kaltenegger, L. 2014 ApJ. 79, 25.
Rodríguez-Mozos, J. M.;Moya, A. 2019, A&A 630, 52.
Sanz-Forcada, J., Micela, G., Ribas, I., et al. 2011 A&A 532, 6.

Selsis, F., Kasting, J. F., Levrard, B., et al. 2007, A&A 476, 137.
Shields A. L., Ballard S. and Johnson J. A. 2016 Phys. Rev. 663, 1.
Shields, A. L. 2019 ApJ. Sup. 243 30.
Tilley, M.A., Segura, A., Meadows, V., et al., 2019 Astrobiology, 19, 1.
Turbet, M, Leconte J, Selsis F., et al. 2016 A&A 596, A112.
Wandel, A. 2018 ApJ. 858, 1.
Wandel, A. 2023a Nature Comm. 14, 2125.
Wandel, A. 2023b Astron. J. 166, 222.
Wandel, A. 2024a Proc. IAUS 387, article 5. Ed. Landt, H., CUP. ISBN 9781009545860 (pub. 2025).
Wandel, A. 2024b IAUS 393, Planetary Science and Exoplanets in the Era of the James Webb Space Telescope.
Wandel. A. 2025, Exoplanets beyond the Conservative Habitable Zone: II. Occurrence, ApJ., accepted.
Wandel, A. and Gale, J. 2020 IJAsB 19, 126W.
Wandel, A. and Tal-Or, L. 2019, ApJ. Lett. 880, L21.
Wandel, A. and Gale, J. 2024, *Life in Space*, Springer, ISBN 978-3-031-64638-6.
Wolf, E. T., Shields, A. L., Kopparapu, R. K. et al. 2017, ApJ. 837, 107.
Yang, J., N. B. Cowan and D. S. Abbot 2013 ApJ. Lett., 771, L45
Yang, J., Boué, G., Fabrycky, D. C. & Abbot, D. S. 2014 ApJ. Lett. 787, L2.
Zieba, S., Kreidberg, L., Ducrot, E., et.al. 2023, Nature 620, 746–749.